\documentclass[sigconf]{acmart}

\usepackage{multirow}
\usepackage{booktabs}
\usepackage{graphicx}
\usepackage[table]{xcolor}
\usepackage{amsmath}
\usepackage{enumitem}
\usepackage{makecell}

\definecolor{bestgreen}{RGB}{210,239,218}
\definecolor{secondblue}{RGB}{217,235,255}

\newcommand{\best}[1]{%
  \cellcolor{bestgreen}\ensuremath{\mathbf{#1}}%
}
\definecolor{speedupgreen}{RGB}{220,245,220}
\newcommand{\secondbest}[1]{%
  \cellcolor{secondblue}\ensuremath{#1}%
}

\newcommand{\qwenbest}[1]{%
  \ensuremath{\underline{#1}}%
}

\newcommand{\softline}{%
  \noalign{\vskip 2pt}%
  \arrayrulecolor{gray!55}%
  \cline{3-13}%
  \arrayrulecolor{black}%
  \noalign{\vskip 2pt}%
}

\AtBeginDocument{%
}

\setcopyright{acmlicensed}
\copyrightyear{2026}
\acmYear{2026}
\acmDOI{XXXXXXX.XXXXXXX}
\acmConference[Conference acronym 'XX]{Make sure to enter the correct
  conference title from your rights confirmation email}{June 03--05,
  2018}{Woodstock, NY}
\acmISBN{978-1-4503-XXXX-X/2018/06}

\newcommand{\blueurl}[1]{\href{#1}{\textcolor{blue}{\nolinkurl{#1}}}}
\begin{document}

\raggedbottom
\title{REPREC: Representation Driven Parameter-Efficient Recommendation System}

\author{Harshini Kavuru}
\email{kavuru.7@osu.edu}
\orcid{1234-5678-9012}
\affiliation{%
  \institution{The Ohio State University}
  \city{Columbus}
  \state{Ohio}
  \country{USA}
}

\author{Dwipam Katariya}
\email{dwipam.katariya@capitalone.com}
\orcid{0009-0009-1058-1244}
\affiliation{%
  \institution{Capital One, AI Foundations}
  \city{Virginia}
  \country{USA}}

\author{Giri Iyengar}
\affiliation{%
  \institution{Capital One, AI Foundations}
  \city{McLean}
  \state{VA}
  \country{USA}
}

\author{Pranab Mohanty}
\affiliation{%
  \institution{Capital One, AI Foundations}
  \city{McLean}
  \state{VA}
  \country{USA}
}

\author{Kalanand Mishra}
\orcid{0000-0002-1832-1537}
\affiliation{%
  \institution{Capital One, AI Foundations}
  \city{San Jose}
  \state{CA}
  \country{USA}
}

\author{Raghu Machiraju}
\email{machiraju.1@osu.edu}
\affiliation{%
  \institution{The Ohio State University}
  \city{Columbus}
  \state{Ohio}
  \country{USA}
}


\begin{abstract}
Large language models (LLMs) have been applied to sequential recommendation by incorporating collaborative signals through input conditioning or model adaptation. However, existing approaches often require LLM fine-tuning, additional architectural modules, representation distillation, or item-level conditioning over long interaction histories, increasing computational and deployment costs. We propose REPREC, a lightweight framework that conditions a frozen LLM using compact user-level representations. REPREC maps a fixed-size embedding from a frozen sequential encoder into a small set of learned soft tokens through an MLP injector, training only the injector while leaving both pretrained backbones unchanged. Our extensive experiments demonstrate that REPREC consistently improves recommendation performance across different sequential encoders, LLM backbones, and user activity levels. Its compact conditioning mechanism also makes REPREC computationally efficient during both training and inference. Moreover, training with short histories while evaluating with longer contexts retains 94--99\% of full-history performance while achieving an average $1.50\times$ per-epoch training speedup. 
\textit{The code is available at} \blueurl{https://github.com/phdbotcode/REPREC}
\end{abstract}



\maketitle

\begin{figure*}[t]
  \centering
  \includegraphics[width=0.9\textwidth]{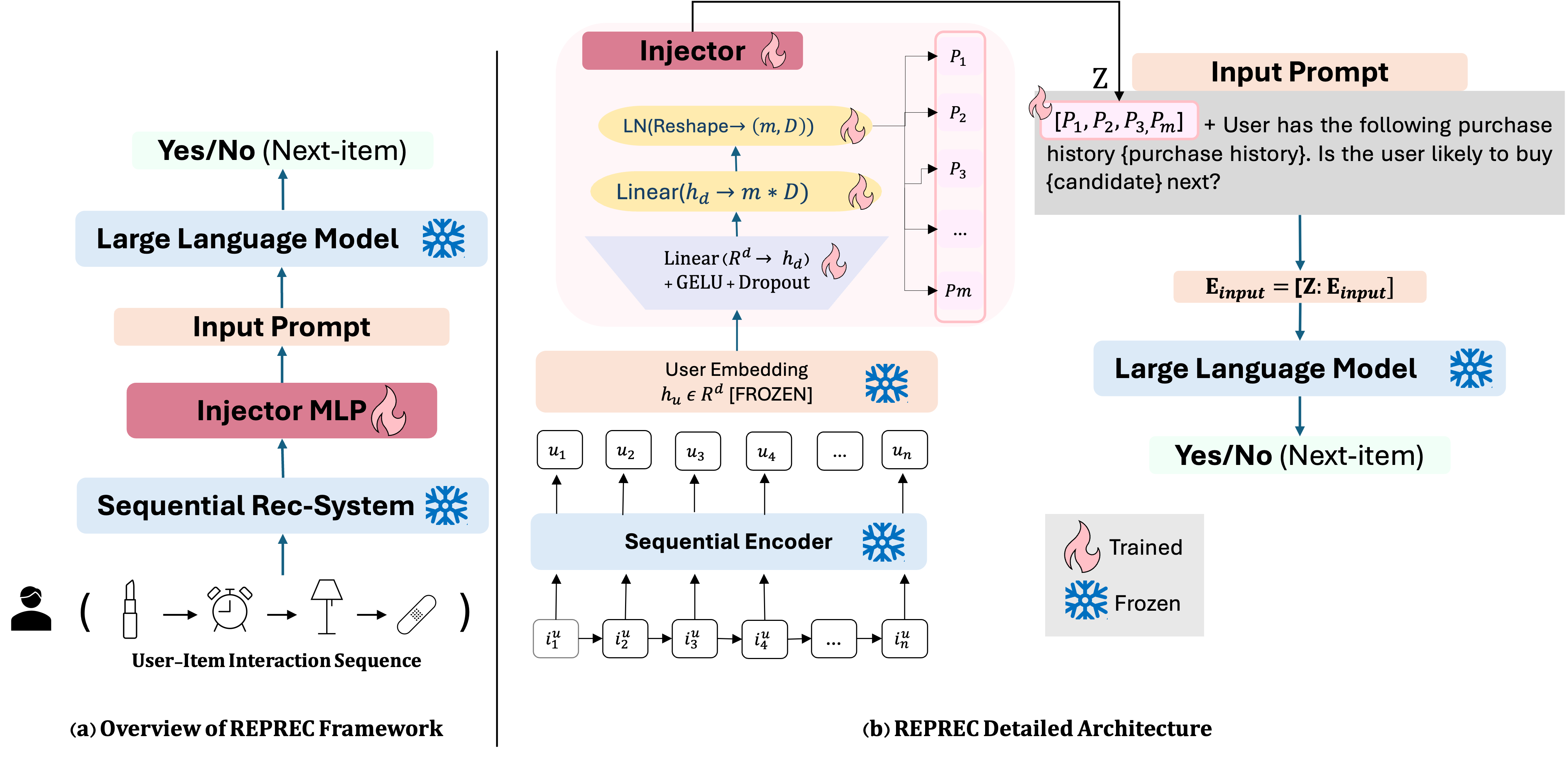}
  \caption{Overview of the REPREC architecture. The sequential
  encoder produces a user embedding projected into $m$ soft
  tokens that condition the frozen LLM for recommendation.}
  \label{fig:overview}
  \Description{Architecture of REPREC showing the sequential
  encoder, MLP injector, and frozen LLaMA backbone.}
\end{figure*}

\section{Introduction}
Sequential recommendation focuses on modeling the order and context of user interactions to predict future behavior. Classical approaches such as collaborative filtering~\cite{10.1145/3038912.3052569, 10.1145/2505515.2505690,CF} and sequential models~\cite{8594844,bert4rec,CLS4Rec} have been highly effective in learning compact user representations that capture behavioral patterns and support personalized ranking. More recently, large language models (LLMs) have been adapted for recommendation by reformulating item prediction as a natural language task~\cite{10.1145/3523227.3546767,10.1145/3708882,LLM4rec,TALLRec}. These approaches show that pretrained LLMs can leverage world knowledge and instruction-following capabilities to perform competitive recommendation without relying on specialized architectures. 

However, textual interaction histories alone may not preserve the structured collaborative signals captured by conventional recommenders. Existing work therefore augments LLMs with representations learned by pretrained collaborative or sequential models. Item-level methods preserve fine-grained behavioral information by conditioning the LLM on representations of individual interactions. For example, LLaRA~\cite{10.1145/3626772.3657690} maps each sequential item embedding to a behavioral token through a learned projection, while
A-LLMRec~\cite{10.1145/3637528.3671931} incorporates collaborative representations of users and items into the LLM input. Although this fine-grained conditioning can improve recommendation quality, injecting one representation per interaction causes the input sequence length to grow with the user's history, increasing attention FLOPs and, consequently, training and inference latency. To avoid representing every historical item independently, subsequent methods move toward user-level integration, but typically require stronger coupling between the recommender and the LLM. CoLLM~\cite{10882951} combines user-level collaborative conditioning with a multi-stage training pipeline that includes LoRA-based LLM adaptation. User-LLM~\cite{10.1145/3701716.3715463} enables richer interaction between recommendation and language representations through dedicated cross-attention modules and joint optimization of the recommendation encoder and LLM. LLM-SRec~\cite{10.1145/3711896.3737035} keeps the LLM frozen but transfers sequential knowledge through a representation distillation objective and additional projection modules. Thus, while
user-level methods reduce dependence on the raw history length, they often introduce backbone adaptation, joint optimization, complex architectures, or encoder-specific alignment procedures. Although these approaches achieve strong recommendation performance, they introduce trade-offs that can limit practical deployment. Most existing methods require modifying or fine-tuning the LLM backbone, jointly optimizing multiple pretrained components, introduce additional architectural modules, or conditioning on item-level representations whose computational cost grows with interaction history length. Consequently, these gains often come at the cost of increased training complexity, larger parameter footprints, reduced modularity, and more difficult integration and maintenance. In real-world recommendation systems, however, model quality is only one consideration: inference latency, training speed, scalability, serving cost, and ease of deployment are equally important, as highlighted by prior industry-focused systems~\cite{alibaba,fintrec,pinnerformer,spotify,spotify2}.

These observations motivate a different design objective: rather than maximizing recommendation performance through increasingly sophisticated adaptation mechanisms, we investigate whether a frozen sequential encoder and a frozen LLM can be connected solely through lightweight user-level representation alignment, while preserving modularity, supporting encoder-agnostic integration, and minimizing deployment complexity. The mechanism of prepending a small number of trainable continuous vectors to condition a frozen language model builds directly on prefix-tuning~\cite{li-liang-2021-prefix} and prompt-tuning~\cite{lester-etal-2021-power}, which learn continuous prompt vectors for a specific task while keeping the underlying LLM frozen. Multimodal approaches such as ClipCap~\cite{clipcap} and BLIP-2~\cite{10.5555/3618408.3619222} extend this idea by conditioning frozen language models on visual soft tokens. Inspired by both lines of work, we study the viability of this fully frozen alignment paradigm for sequential recommendation. To the best of our knowledge, REPREC is the first work to systematically investigate whether a frozen sequential recommender and a frozen LLM can be connected using only a trainable user-level representation injector, without adapting either pretrained backbone. We introduce \textbf{REPREC} (\textbf{Rep}resentation-driven \textbf{P}arameter-\textbf{E}fficient \textbf{Rec}ommendation \textbf{S}ystem), a lightweight framework that maps a user representation produced by a frozen sequential encoder into a small set of soft conditioning tokens
prepended to a frozen LLM. All adaptation is confined to a compact MLP injector, avoiding modifications to either backbone. We investigate the following research questions:
\begin{itemize}[topsep=0pt, itemsep=0pt, parsep=0pt, partopsep=0pt]
\item \noindent\textbf{RQ1:} Can soft token injection align user representations with the LLM input space for effective recommendation without modifying the backbone?
\item \noindent\textbf{RQ2:} How does performance vary across casual, core, and power user groups?
\item \noindent\textbf{RQ3:} Can REPREC trained on short prompt histories generalize to longer histories at inference time?
\end{itemize}

\section{Methodology}
\subsection{Problem Definition}

Let $\mathcal{D}$ denote a historical interaction dataset consisting of triplets $(u, i, y)$, where $u \in \mathcal{U}$ is a user, $i \in \mathcal{V}$ is an item, and $y \in \{0,1\}$ indicates whether the interaction occurred. Each user $u$ is associated with a chronological interaction sequence:\begin{equation}
S_u = (i_1^{(u)}, i_2^{(u)}, \dots, i_{n_u}^{(u)}),
\end{equation}

where $i_t^{(u)} \in \mathcal{V}$ and $n_u$ denotes the number of interactions of user $u$. We adopt the standard next-item recommendation setting. Given a user $u$ and their interaction prefix, the objective is to predict the next item $i_{t+1}^{(u)}$ among candidate items from $\mathcal{V}$. We adopt evaluation strategy of sasrec \cite{8594844}, specifically the model ranks the ground-truth next item against sampled negative items (see Appendix~\ref{app:negsampling} for the sampling procedure). Performance is measured using HIT@K.

\subsection{REPREC}
REPREC consists of three components as illustrated in
Figure~\ref{fig:overview}:
(i) a sequential encoder $f_{\theta}$ (the sequential encoder is modular and can be replaced with any recommendation backbones),
(ii) a lightweight injector $g_{\phi}$ that maps structured user
embeddings into the LLM embedding space, and
(iii) a frozen LLM backbone $\mathcal{M}$.

\subsubsection{Sequential Encoder:}

We first learn structured behavioral representations using a sequential recommender model. Given a user interaction prefix
$S_u^{\le t} = (i_1^{(u)}, \dots, i_t^{(u)})$
we train a sequential encoder
\[
f_\theta : \mathcal{V}^* \rightarrow \mathbb{R}^d
\]
that maps the sequence of item IDs into a $d$-dimensional embedding. 
\[
\mathbf{h}_u = f_\theta(S_u^{\le t}), 
\qquad \mathbf{h}_u \in \mathbb{R}^d.
\]
The encoder is trained using standard next-item prediction, where prediction is computed via a dot-product between the user representation and candidate item embeddings. After training converges, $f_\theta$ is frozen. In this work, we instantiate $f_\theta$ as SASRec~\cite{8594844} and BERT4Rec~\cite{bert4rec}.

\begin{table*}[!t]
\centering
\caption{
Performance comparison across five datasets.
Green and light-blue cells indicate the best and second-best
performance, respectively, in each metric column.
Underlined values indicate the best mean performance among
methods using Qwen as the LLM. $^{**}$ indicates REPREC result significantly outperforms all baseline methods (paired $t$-test over per-user HIT@$K$ indicators, $p<0.005$).
}
\label{tab:main_results}

\scriptsize
\setlength{\tabcolsep}{2.8pt}
\renewcommand{\arraystretch}{1.08}

\resizebox{\textwidth}{!}{%
\begin{tabular}{lll||c|c||c|c||c|c||c|c||c|c}
\toprule

\multirow{2}{*}{\textbf{Setting}}
& \multirow{2}{*}{\textbf{Backbone}}
& \multirow{2}{*}{\textbf{Method}}
& \multicolumn{2}{c||}{\textbf{Beauty}$^{**}$}
& \multicolumn{2}{c||}{\textbf{Sports}$^{**}$}
& \multicolumn{2}{c||}{\textbf{Toys \& Games}}
& \multicolumn{2}{c||}{\textbf{Pet Supplies}}
& \multicolumn{2}{c}{\textbf{Tools \& Home}} \\

\cmidrule(lr){4-5}
\cmidrule(lr){6-7}
\cmidrule(lr){8-9}
\cmidrule(lr){10-11}
\cmidrule(lr){12-13}

& &
& \textbf{HIT@5}
& \textbf{HIT@10}
& \textbf{HIT@5}
& \textbf{HIT@10}
& \textbf{HIT@5}
& \textbf{HIT@10}
& \textbf{HIT@5}
& \textbf{HIT@10}
& \textbf{HIT@5}
& \textbf{HIT@10} \\

\midrule


\multirow{2}{*}{Sequential}
& \multirow{2}{*}{--}
& SASRec
& $0.172$
& $0.247$
& $0.130$
& $0.202$
& $0.177$
& $0.246$
& $0.126$
& $0.204$
& $0.097$
& $0.153$ \\

& &
BERT4Rec
& $0.080$
& $0.141$
& $0.026$
& $0.054$
& $0.063$
& $0.116$
& $0.081$
& $0.153$
& $0.045$
& $0.085$ \\

\midrule


\multirow{6}{*}{Baseline}
& \multirow{5}{*}{LLaMA}
& Zero-shot
& $0.070$
& $0.105$
& $0.050$
& $0.090$
& $0.050$
& $0.090$
& $0.051$
& $0.089$
& $0.053$
& $0.085$ \\

& &
Frozen Injector
& $0.065$
& $0.101$
& $0.040$
& $0.072$
& $0.040$
& $0.072$
& $0.042$
& $0.073$
& $0.052$
& $0.083$ \\

& &
Prompt Tuning
& $0.298$
& $0.394$
& $0.255$
& $0.366$
& $0.315$
& $0.386$
& $0.245$
& $0.345$
& $0.201$
& $0.283$ \\

\softline

& &
TALLRec/LoRA-FT
& $0.327$
& $0.424$
& $0.248$
& $0.365$
& \best{0.346}
& $0.421$
& $0.264$
& $0.368$
& $0.233$
& \secondbest{0.319} \\

\softline

& &
LLaRA-S
& $0.182$
& $0.328$
& $0.169$
& $0.250$
& $0.310$
& $0.395$
& $0.230$
& $0.327$
& $0.190$
& $0.267$ \\

& &
LLaRA-B
& $0.159$
& $0.281$
& $0.095$
& $0.103$
& \best{0.346}
& \secondbest{0.423}
& $0.248$
& $0.359$
& $0.229$
& $0.312$ \\

\softline

& &
CoLLM-S
& $0.302$
& $0.389$
& $0.288$
& \secondbest{0.401}
& $0.326$
& $0.401$
& $0.135$
& $0.221$
& $0.125$
& $0.190$ \\

& &
CoLLM-B
& $0.328$
& $0.426$
& $0.286$
& $0.390$
& $0.327$
& $0.398$
& $0.255$
& $0.357$
& $0.228$
& $0.314$ \\

\cmidrule(lr){2-13}

&
Qwen
& TALLRec/LoRA-FT
& $0.317$
& $0.412$
& $0.251$
& $0.360$
& \qwenbest{0.334}
& $0.408$
& $0.259$
& $0.361$
& $0.221$
& \qwenbest{0.307} \\

\midrule


\multirow{4}{*}{REPREC}
& \multirow{2}{*}{LLaMA}
& REPREC-S
& \best{0.351 \pm 0.010}
& \best{0.450 \pm 0.010}
& \best{0.295 \pm 0.004}
& \best{0.407 \pm 0.006}
& \best{0.346 \pm 0.001}
& \best{0.425 \pm 0.001}
& \secondbest{0.268 \pm 0.004}
& \secondbest{0.375 \pm 0.003}
& \best{0.238 \pm 0.005}
& \secondbest{0.319 \pm 0.003} \\

& &
REPREC-B
& \secondbest{0.346 \pm 0.001}
& \secondbest{0.444 \pm 0.001}
& \secondbest{0.275 \pm 0.002}
& $0.385 \pm 0.000$
& \secondbest{0.339 \pm 0.004}
& $0.415 \pm 0.005$
& \best{0.270 \pm 0.010}
& \best{0.378 \pm 0.012}
& \secondbest{0.237 \pm 0.004}
& \best{0.323 \pm 0.003} \\

\cmidrule(lr){2-13}

&
\multirow{2}{*}{Qwen}
& REPREC-S
& \qwenbest{0.320 \pm 0.002}
& \qwenbest{0.414 \pm 0.004}
& $0.257 \pm 0.006$
& $0.364 \pm 0.007$
& $0.333 \pm 0.007$
& \qwenbest{0.410 \pm 0.005}
& \qwenbest{0.263 \pm 0.004}
& \qwenbest{0.367 \pm 0.001}
& $0.222 \pm 0.006$
& $0.303 \pm 0.008$ \\

& &
REPREC-B
& $0.316 \pm 0.002$
& \qwenbest{0.414 \pm 0.004}
& \qwenbest{0.265 \pm 0.007}
& \qwenbest{0.370 \pm 0.010}
& $0.317 \pm 0.001$
& $0.394 \pm 0.002$
& $0.256 \pm 0.004$
& $0.361 \pm 0.001$
& \qwenbest{0.226 \pm 0.003}
& $0.304 \pm 0.002$ \\

\bottomrule
\end{tabular}%
}
\end{table*}

\begin{figure*}[!t]
\centering
\includegraphics[width=0.9\textwidth]{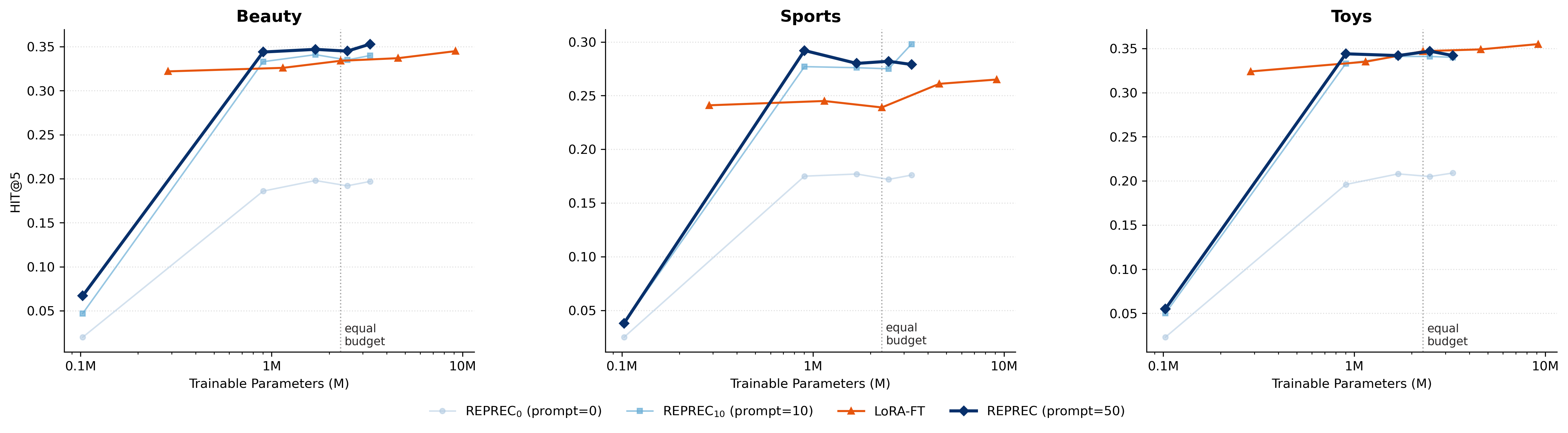}
\caption{
HIT@5 versus the number of trainable parameters using LLaMA as LLM. For REPREC, we vary the number of soft tokens as $m \in \{2,4,6,8\}$, while for LoRA we vary the rank as $r \in \{4,8,16,32\}$. Each marker corresponds to one value of $m$ for REPREC or one value of $r$ for LoRA-FT. The dashed vertical line indicates the matched trainable-parameter budget. REPREC$_0$, REPREC$_{10}$, and REPREC denote training and inference with 0, 10, and 50 recent interaction-history items in the textual prompt, respectively, while using the same learned user-representation tokens.
}
\label{fig:param_efficiency}
\end{figure*}

\subsubsection{Injector: User Representation-to-LLM Projection}

We introduce a projection function $g_\phi : \mathbb{R}^d \rightarrow \mathbb{R}^{m \times D}$, parameterized by $\phi$:

\begin{itemize}
    \item $d$ is the sequential embedding dimension,
    \item $D$ is the LLM hidden size,
    \item $m$ is the number of injected soft tokens.
\end{itemize}

The injector is implemented as a multi-layer perceptron (MLP):
\[
g_\phi(\mathbf{h}_u) =
\mathrm{LN}_{D}\!\left(
  \operatorname{reshape}_{m \times D}\!\left(
    \mathbf{W}_2\,\sigma(\mathbf{W}_1\mathbf{h}_u + \mathbf{b}_1) + \mathbf{b}_2
  \right)
\right),
\]
where $\sigma$ is a non-linear activation (GELU in our implementation).
The projected output is reshaped into $m$ token embeddings of 
dimension $D$, then normalized via a single shared 
$\mathrm{LayerNorm}$ applied independently over the $D$-dimensional 
embedding of each token before being prepended to the LLM input space. The output $\mathbf{Z} = g_\phi(\mathbf{h}_u) \in \mathbb{R}^{m \times D}$ constitutes $m$ soft conditioning tokens in the LLM embedding space.

\subsubsection{Prompt Construction and Training:} Given a candidate item $i$, we construct a binary decision
$\text{prompt}(i)$ and prepend the injected user tokens $\mathbf{Z}$. $\mathbf{E}_{input} = [\mathbf{Z}; \mathbf{E}_{prompt}]$, where $[\cdot;\cdot]$ denotes sequence-dimension concatenation. The frozen LLM $\mathcal{M}$ produces vocabulary logits over $\mathbf{E}_{input}$, and we extract the probability of the target answer token (\texttt{Yes}/\texttt{No}):
\[
p(y=1 \mid u, i) = \text{softmax}(\mathbf{W}_{llm} \mathbf{h}_{T})_{\text{Yes}},
\]
where $\mathbf{h}_{T} \in \mathbb{R}^D$ is the hidden state of
$\mathcal{M}$ at the final token position $T$.
We compute cross-entropy loss only at the answer position,
masking all prompt tokens:
\[
\mathcal{L}(\phi)
= \mathbb{E}_{(u,i,y) \sim \mathcal{D}}
\left[- \log p_\phi(y \mid u, i)\right],
\]
where gradients flow through LLM activations into the injector
parameters $\phi$ while the backbone remains frozen.

\begin{table}[h]
\centering
\small
\caption{Statistics of datasets after preprocessing.
         Avg.\ Len denotes the average sequence length per user.
         Casual, Core, and Power report the percentage of users
         in each activity group.}
\label{tab:dataset_stats}

\resizebox{\columnwidth}{!}{%
\begin{tabular}{lrrrrrrr}
\toprule
\textbf{Dataset}
  & \textbf{\#Users}
  & \textbf{\#Items}
  & \textbf{\#Int.}
  & \textbf{Avg.\ Len}
  & \textbf{Casual}
  & \textbf{Core}
  & \textbf{Power} \\
\midrule

Beauty
  & 22,363
  & 12,101
  & 198,498
  & 8.88
  & 62.9\%
  & 32.9\%
  & 4.2\% \\

Sports \& Outdoors
  & 35,598
  & 18,357
  & 296,241
  & 8.32
  & 63.6\%
  & 33.6\%
  & 2.8\% \\

Toys \& Games
  & 19,412
  & 11,924
  & 167,247
  & 8.62
  & 65.3\%
  & 30.9\%
  & 3.8\% \\

Pet Supplies
  & 19,856
  & 8,510
  & 157,836
  & 7.95
  & 66.1\%
  & 31.8\%
  & 2.1\% \\

Tools \& Home Improvement
  & 16,638
  & 10,217
  & 134,476
  & 8.08
  & 66.0\%
  & 31.5\%
  & 2.5\% \\

\bottomrule
\end{tabular}%
}
\end{table}

\begin{table*}[!t]
\centering
\caption{
HIT@10 performance across user activity groups.
Green and light-blue cells indicate the best and second-best performance, respectively, within each dataset and user activity group.
}
\label{tab:user_groups}

\scriptsize
\setlength{\tabcolsep}{3.2pt}
\renewcommand{\arraystretch}{1.10}

\resizebox{\textwidth}{!}{%
\begin{tabular}{l|ccc|ccc|ccc|ccc|ccc}
\toprule

\multirow{2}{*}{\textbf{Method}}
& \multicolumn{3}{c|}{\textbf{Beauty}}
& \multicolumn{3}{c|}{\textbf{Sports}}
& \multicolumn{3}{c|}{\textbf{Toys \& Games}}
& \multicolumn{3}{c|}{\textbf{Tools \& Home}}
& \multicolumn{3}{c}{\textbf{Pet Supplies}} \\

\cmidrule(lr){2-4}
\cmidrule(lr){5-7}
\cmidrule(lr){8-10}
\cmidrule(lr){11-13}
\cmidrule(lr){14-16}

& \textbf{Casual} & \textbf{Core} & \textbf{Power}
& \textbf{Casual} & \textbf{Core} & \textbf{Power}
& \textbf{Casual} & \textbf{Core} & \textbf{Power}
& \textbf{Casual} & \textbf{Core} & \textbf{Power}
& \textbf{Casual} & \textbf{Core} & \textbf{Power} \\

\midrule

SASRec
& 0.227 & 0.268 & 0.383
& 0.204 & 0.197 & 0.193
& 0.239 & 0.252 & 0.330
& 0.154 & 0.151 & 0.134
& 0.208 & 0.195 & 0.205 \\

TALLRec
& 0.408 & 0.412 & 0.442
& 0.359 & 0.341 & 0.323
& 0.423 & 0.416 & 0.353
& 0.313 & \secondbest{0.301} & 0.263
& 0.366 & 0.360 & 0.371 \\

LLaRA
& 0.236 & 0.311 & 0.444
& 0.239 & 0.248 & 0.305
& 0.376 & 0.405 & \best{0.420}
& 0.245 & 0.257 & \secondbest{0.277}
& 0.319 & 0.321 & 0.395 \\

CoLLM
& 0.410 & 0.414 & 0.457
& 0.383 & 0.372 & \secondbest{0.354}
& 0.396 & 0.397 & 0.331
& 0.308 & 0.294 & \secondbest{0.277}
& 0.354 & 0.350 & 0.409 \\

\midrule

REPREC
& \secondbest{0.428} & \secondbest{0.437} & \best{0.494}
& \secondbest{0.396} & \best{0.376} & \best{0.359}
& \secondbest{0.425} & \best{0.417} & \secondbest{0.397}
& \secondbest{0.318} & \secondbest{0.301} & \best{0.311}
& \secondbest{0.376} & \secondbest{0.370} & \best{0.446} \\

REPREC$_{10}$
& \best{0.429} & \best{0.449} & \secondbest{0.470}
& \best{0.396} & \secondbest{0.374} & 0.348
& \best{0.426} & \best{0.417} & 0.373
& \best{0.319} & \best{0.306} & \secondbest{0.277}
& \best{0.377} & \best{0.376} & \secondbest{0.417} \\

\bottomrule
\end{tabular}%
}

\end{table*}

\begin{table}[t]
\centering
\caption{
Efficiency comparison across five datasets.
All runtimes are reported in minutes, with training time measured per epoch,
and all experiments are conducted on a single NVIDIA H100 GPU under the same
hardware setting. Bold values indicate the lowest inference time for each
dataset. Speedup reports the average inference-time speedup of REPREC over
each baseline across datasets.
}
\label{tab:runtime}

\scriptsize
\setlength{\tabcolsep}{3.5pt}
\renewcommand{\arraystretch}{1.08}

\resizebox{\columnwidth}{!}{%
\begin{tabular}{lc|c|c|c||c}
\toprule

\textbf{Dataset}
& \textbf{Time}
& \textbf{TALLRec}
& \textbf{LLaRA}
& \textbf{CoLLM}
& \textbf{REPREC} \\

\midrule

\multirow{2}{*}{Beauty}
& Training  & 357.97 & 796.07 & 665.48 & \textbf{278.12} \\
& Inference & 375.97 & 683.21 & 421.42 & \textbf{348.43} \\

\midrule

\multirow{2}{*}{Sports}
& Training  & 213.63 & 569.56 & 413.40 & \textbf{186.43} \\
& Inference & 412.64 & 814.23 & 473.81 & \textbf{385.94} \\

\midrule

\multirow{2}{*}{Toys \& Games}
& Training  & 133.39 & 356.90 & 267.09 & \textbf{114.96} \\
& Inference & 179.51 & 363.84 & 207.37 & \textbf{173.35} \\

\midrule

\multirow{2}{*}{Pet Supplies}
& Training  & 116.57 & 294.16 & 232.79 & \textbf{99.91} \\
& Inference & 219.57 & 431.61 & 250.83 & \textbf{205.22} \\

\midrule

\multirow{2}{*}{Tools \& Home}
& Training  & 148.66 & 372.06 & 300.66 & \textbf{130.35} \\
& Inference & 232.64 & 447.27 & 265.35 & \textbf{217.52} \\

\midrule
\midrule


\multicolumn{2}{l|}{\textbf{Trainable Params.}}
& 2.29M
& 12.16M
& 4.35M
& \textbf{2.39M} \\

\midrule


\multirow{2}{*}{\textbf{REPREC Speedup}}
& \cellcolor{speedupgreen}\textbf{Training}
& \cellcolor{speedupgreen}\textbf{1.18$\times$}
& \cellcolor{speedupgreen}\textbf{2.96$\times$}
& \cellcolor{speedupgreen}\textbf{2.31$\times$}
& \cellcolor{speedupgreen}-- \\

& \cellcolor{speedupgreen}\textbf{Inference}
& \cellcolor{speedupgreen}\textbf{1.06$\times$}
& \cellcolor{speedupgreen}\textbf{2.07$\times$}
& \cellcolor{speedupgreen}\textbf{1.22$\times$}
& \cellcolor{speedupgreen}-- \\

\bottomrule
\end{tabular}%
}
\end{table}

\section{Experiments}
\subsection{Data}
\noindent We evaluate on five Amazon product review benchmarks from the
2018 release~\cite{Amazon-data}. For each user, the last interaction is held out for testing, the second-to-last for validation, and the remainder for
training. All datasets are highly sparse, with average sequence lengths
between 7.95 and 8.88 interactions per user. Table~\ref{tab:dataset_stats} summarizes the statistics after preprocessing.

\subsection{Baselines}
We compare REPREC against two groups of baselines: conventional
sequential recommendation models and LLM-based recommendation
methods. For the sequential recommendation baselines, we evaluate
SASRec~\cite{8594844} and BERT4Rec~\cite{bert4rec} as standalone
models. These models also serve as the pretrained sequential
encoders used by REPREC. For the LLM-based baselines, we consider
Zero-shot LLM, Frozen Injector, Prompt Tuning, LoRA-FT~\cite{hu2022lora}, TALLRec~\cite{TALLRec},
LLaRA~\cite{10.1145/3626772.3657690}, and CoLLM~\cite{10882951}. The
Zero-shot LLM provides a frozen LLM with the user's recent
interactions represented as text, without parameter adaptation or
injected collaborative information. Frozen Injector prepends soft
tokens generated by a randomly initialized, untrained projection
module while keeping the LLM frozen, thereby controlling for
improvements caused solely by introducing continuous prompt tokens. Prompt Tuning replaces REPREC's injector with
$m{=}6$ directly trained soft-token embeddings, shared across all
users and independent of any sequential encoder, thereby controlling
for improvements attributable to trained prefix capacity alone,
independent of user-specific conditioning.
LoRA-FT adapts the LLM through low-rank fine-tuning using the
interaction history in textual form. TALLRec is not
evaluated as a separate baseline; its LoRA configuration and text-only
conditioning reduce to the same setup as LoRA-FT (see
Appendix~D.1 for the detailed justification). LLaRA projects
item-level representations from a pretrained sequential recommender
into the LLM token space; following its original configuration, we
use \textbf{S}ASRec(LLaRA-\textbf{S}) and \textbf{B}ERT4Rec(LLaRA-\textbf{B})
as its sequential encoders. CoLLM similarly conditions the LLM on a
pretrained sequential recommender, but via a two-stage recipe:
LoRA is first fine-tuned on text-only prompts, then frozen while a
projector is trained to map collaborative user and item embeddings
into the LLM's token space; we likewise use
\textbf{S}ASRec(CoLLM-\textbf{S}) and \textbf{B}ERT4Rec(CoLLM-\textbf{B})
as its sequential encoders. We evaluate all six LLM-based baselines
using LLaMA as the backbone. To examine generalization across LLM
backbones, we additionally evaluate Qwen, using LoRA-FT as the
adapted-LLM baseline for comparison with REPREC. REPREC is evaluated
with both \textbf{S}ASRec(REPREC-\textbf{S}) and
\textbf{B}ERT4Rec(REPREC-\textbf{B}) as sequential encoders under
the LLaMA and Qwen backbones.

\subsection{Implementation Details}
We use a maximum interaction-history length of 50 for all models and evaluate recommendation performance using HIT@5 and HIT@10. For the LLM-based methods, recommendation is formulated as a binary candidate-relevance task. Candidate items are ranked using the model's relevance scores, and Hit Rate is computed from the resulting ranking. To ensure comparison under matched trainable-parameter budgets, we use LoRA rank $r{=}8$ for LoRA-FT/TALLRec, LLaRA, and CoLLM, and $m{=}6$ soft tokens for REPREC; LLaRA and CoLLM additionally train their respective projection modules. Additional parameter-budget details are provided in Appendix~A. The results for REPREC are averaged over five random seeds and reported as mean and standard deviation to assess robustness to variations in model initialization.

\subsection{Performance and Efficiency Comparison}
\subsubsection{\textbf{Overall Performance (RQ1):}}
Table~\ref{tab:main_results} reports the overall performance comparison across all baselines. The Frozen Injector consistently performs worse than the zero-shot baseline, confirming that simply prepending untrained embeddings is insufficient; the collaborative representation must be explicitly aligned with the LLM embedding space to provide a useful conditioning signal. Prompt Tuning, which
trains the same number of soft tokens without any user-specific
input, substantially improves over both Zero-shot and Frozen
Injector, confirming that trainable prefix capacity itself provides
a meaningful conditioning signal; however, it remains below REPREC on every dataset, indicating that trained capacity alone is insufficient and that user-specific conditioning
provides additional benefit. Consistent with this observation, Figure~\ref{fig:param_efficiency} shows that injecting as few as two learned soft tokens already yields substantial improvements over the zero-shot baseline, indicating that a compact learned user representation is sufficient to effectively condition the frozen LLM. Compared with standalone sequential recommenders, REPREC consistently outperforms SASRec and BERT4Rec, demonstrating the benefit of combining collaborative user representations with the semantic knowledge encoded by the pretrained LLM. REPREC also achieves the strongest or competitive performance against LLM-based baselines under matched trainable-parameter budgets (see Appendix~\ref{app:params} for details), including TALLRec, LLaRA, and CoLLM, while keeping the LLM backbone frozen.

Figure~\ref{fig:param_efficiency} further shows that soft-token conditioning alone is insufficient. Using no textual interaction history (prompt=0) consistently underperforms configurations that include recent interaction histories (prompt=10 and prompt=50). As additional prompt history is incorporated, REPREC achieves progressively better recommendation performance, suggesting that the learned user representation and textual interaction history provide complementary information. Finally, both REPREC and LoRA exhibit diminishing returns as model capacity increases, with only marginal gains beyond $m{=}2$ soft tokens for REPREC and $r{=}8$ for LoRA, suggesting that additional trainable parameters provide limited benefit in this setting.

\subsubsection{\textbf{Training and Inference Efficiency:}}

Beyond recommendation performance, we evaluate the computational efficiency of REPREC against existing LLM-based recommendation approaches. Table~\ref{tab:runtime} reports the training and inference time across all five datasets. REPREC is consistently efficient because it keeps both the LLM and sequential recommender frozen and trains only the lightweight representation injector. For a controlled comparison, we use LoRA rank $r=8$ and $m=6$ soft tokens for REPREC to maintain comparable trainable-parameter budgets; despite this, REPREC exhibits lower training and inference runtime. LLaRA incurs additional computational cost by projecting collaborative item embeddings into the LLM input, causing the number of behavioral tokens to grow with the interaction history. Under the standard Transformer compute approximation of approximately $6NT$, where $N$ denotes the number of model parameters and $T$ the number of processed tokens, increasing the input sequence length increases computational cost and consequently training and inference latency. In contrast, REPREC compresses the complete collaborative history into a fixed set of $m$ user-level soft tokens, preventing this history-dependent growth in LLM input length. CoLLM further introduces a multi-stage training pipeline, first adapting the LLM with LoRA ($r=8$) and subsequently incorporating collaborative user and target-item representations in a second training stage, resulting in additional training overhead compared with REPREC's single-stage representation alignment. Overall, these results highlight the favorable performance-efficiency trade-off of REPREC; rather than fine-tuning the LLM, increasing the prompt length with item-level collaborative tokens, or requiring multiple training stages, REPREC uses a compact, fixed-length representation to condition a frozen LLM, providing a simple and computationally efficient approach while maintaining strong recommendation performance.

\begin{table*}[t]
\centering
\small
\setlength{\tabcolsep}{4pt}

\caption{
Performance and efficiency evaluated for the Power User persona.
REPREC$_{\text{cheap}}$ (train $\ell{=}10$, test $\ell{=}50$) is compared
with REPREC ($\ell{=}50$). Performance ratio denotes the HIT@10 retained
by REPREC$_{\text{cheap}}$ relative to REPREC.
}
\label{tab:efficiency}

\begin{tabular}{l|cc|cc|c|cc}
\toprule

\multirow{2}{*}{\textbf{Dataset}}
& \multicolumn{2}{c|}{\textbf{HIT@10}}
& \multicolumn{2}{c|}{\textbf{Train Time (min/epoch)}}
& \multicolumn{1}{c|}{\textbf{Performance Ratio}}
& \multicolumn{2}{c}{\textbf{Efficiency}} \\

\cmidrule(lr){2-3}
\cmidrule(lr){4-5}
\cmidrule(lr){6-6}
\cmidrule(lr){7-8}

& \textbf{REPREC}
& \textbf{REPREC$_{\text{cheap}}$}
& \textbf{REPREC}
& \textbf{REPREC$_{\text{cheap}}$}
& \textbf{REPREC$_{\text{cheap}}$/REPREC}
& \textbf{Saved (min)}
& \textbf{Speedup} \\

\midrule

Beauty
& 0.494 & 0.473
& 278.12 & 235.84
& 0.96$\times$
& 42.28 & 1.18$\times$ \\

Sports
& 0.359 & 0.355
& 186.43 & 116.28
& 0.99$\times$
& 70.15 & 1.60$\times$ \\

Toys
& 0.397 & 0.381
& 114.96 & 67.67
& 0.96$\times$
& 47.29 & 1.70$\times$ \\

Pet Supplies
& 0.446 & 0.427
& 99.91 & 71.15
& 0.96$\times$
& 28.77 & 1.40$\times$ \\

Tools \& Home
& 0.311 & 0.292
& 130.35 & 80.02
& 0.94$\times$
& 50.33 & 1.63$\times$ \\

\bottomrule
\end{tabular}
\end{table*}

\subsubsection{\textbf{User Regimes (RQ2):}} 
To understand how different sources of user information affect recommendation across activity levels, we divide users into three personas according to sequence length: Casual users with 0-5 interactions, Core users with 6-20 interactions, and Power users with more than 20 interactions. The proportion of users in each persona is reported in Table~\ref{tab:dataset_stats}. We compare REPREC against baselines within each regime. For LLM-based methods that incorporate a collaborative encoder, we report the encoder backbone that yields the best performance for that method. Across the three user regimes, REPREC achieves the strongest overall performance, demonstrating that its user-level representation remains effective across both sparse and interaction-rich histories.

The comparison across baselines further reveals how the utility of textual and collaborative signals changes with user activity. TALLRec relies exclusively on textual interaction histories and adapts the LLM through LoRA. While it remains competitive for Casual and Core users, its relative performance decreases for Power users compared with LLM-based methods that explicitly incorporate collaborative signals. In contrast, LLaRA, which injects collaborative item representations into the prompt, becomes particularly competitive for Power users, suggesting that item-level collaborative information becomes increasingly useful as richer interaction histories become available. CoLLM exhibits a complementary pattern: incorporating collaborative user and item representations allows it to outperform TALLRec for Power users while also performing more strongly than LLaRA for Casual and Core users. REPREC consistently outperforms these alternatives across user regimes, indicating that compressing collaborative history into a compact user-level representation provides an effective conditioning signal without requiring item-level injection or adaptation of the LLM backbone.

We further investigate the role of textual history through REPREC$_{10}$, which is trained and evaluated using only the 10 most recent interactions in the textual prompt. Interestingly, for Casual and Core users, REPREC$_{10}$ matches or slightly exceeds full-history REPREC across datasets, indicating that a shorter textual context is sufficient when interaction histories are sparse and that the LLM can rely more on its pretrained semantic knowledge together with the injected collaborative representation. This trend reverses for Power users, where REPREC$_{10}$ consistently underperforms REPREC. For users with richer behavioral histories, retaining a longer textual interaction context therefore provides additional information that cannot be fully recovered from a truncated prompt. 

\begin{figure}[t]
    \centering
    \includegraphics[width=\columnwidth]{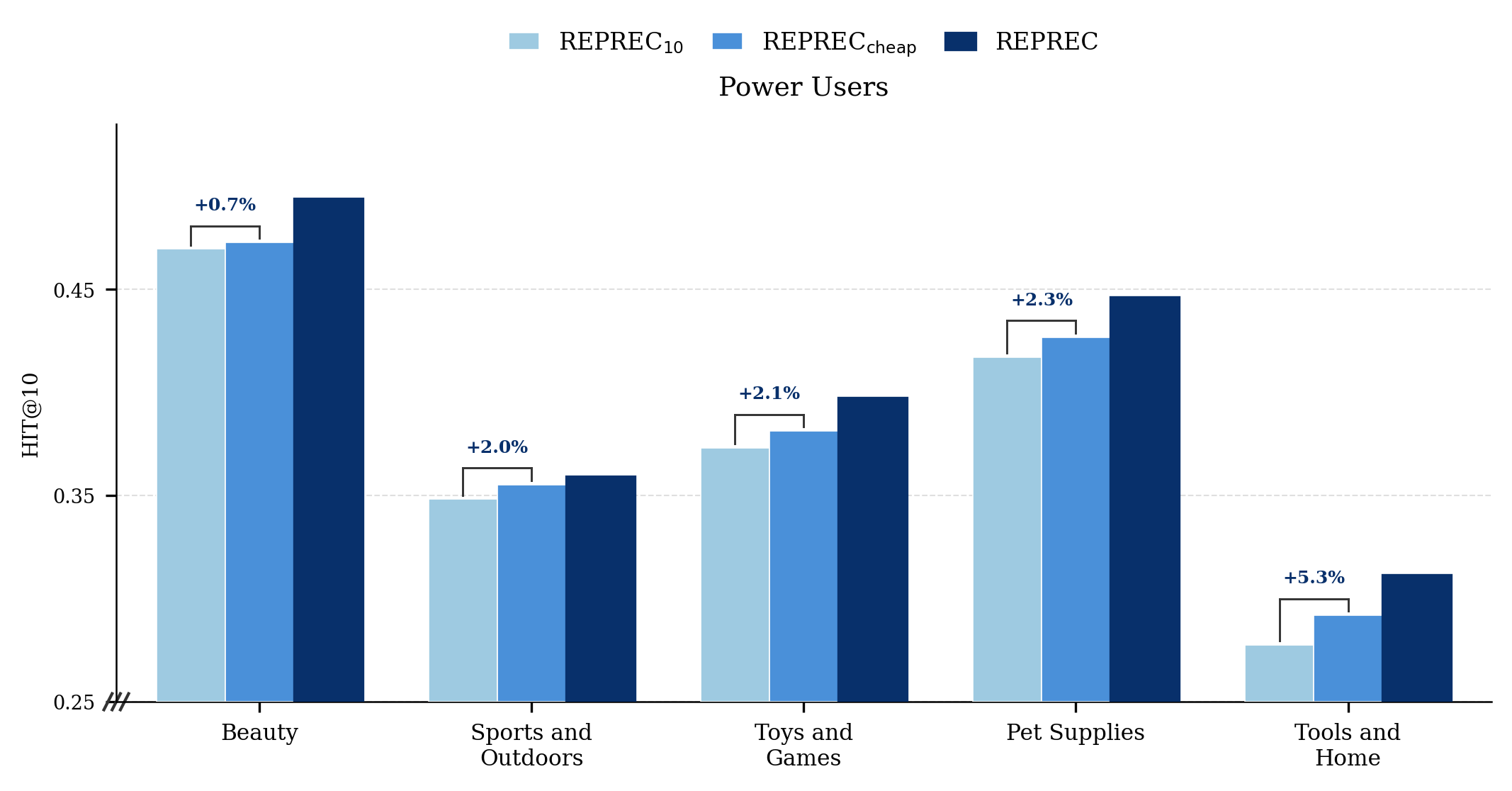}
    \caption{
    HIT@10 performance for power users across five datasets.
    Percentages indicate the relative improvement of
    REPREC$_{\mathrm{cheap}}$ over REPREC$_{10}$.
    }
    \label{fig:power_users}
\end{figure}

\subsubsection{\textbf{Training and Inference Prompt Lengths (RQ3):}}

We study whether REPREC can reduce training cost by using shorter interaction histories without sacrificing the benefits of longer histories at inference. As shown in Figure~\ref{fig:param_efficiency}, training and evaluating with a short history ($\ell=10$) remains competitive at the aggregate level; however, the user-level analysis in Figure~\ref{fig:power_users} shows that this setting degrades performance for power users, who benefit more from longer interaction histories. Motivated by this observation, we introduce REPREC$_{\text{cheap}}$, which is trained with $\ell=10$ but evaluated with $\ell=50$. Figure~\ref{fig:power_users} shows that increasing the inference history from REPREC${10}$ to REPREC${\text{cheap}}$, while keeping the training history fixed, consistently improves performance across all five datasets, yielding relative HIT@10 gains of
0.7\% to 5.3\%. This recovery is achieved without additional training, indicating that an alignment module learned from short histories can effectively exploit richer interaction context when it becomes available at inference. As shown in Table~\ref{tab:efficiency}, REPREC$_{\text{cheap}}$ consequently retains 94 to 99\% of the HIT@10 achieved by full-history REPREC on power users, while reducing training cost by 1.18$\times$--1.70$\times$ per epoch (1.50$\times$ on average). These results demonstrate that REPREC can learn its representation alignment efficiently from short histories while benefiting from longer contexts at inference, providing a favorable trade-off between training efficiency and recommendation performance.

\subsection{Ablation Studies}
\subsubsection{\textbf{Generalization Across LLM Backbones:}}
To assess whether REPREC's effectiveness depends on a particular LLM backbone, we evaluate REPREC-S and REPREC-B with two parameter-matched, instruction-tuned models from different families, LLaMA (\texttt{Llama- 3.2-3B-Instruct}) and Qwen (\texttt{Qwen2.5-3B-Instruct}). REPREC remains effective under both backbones, demonstrating that the proposed representation-alignment mechanism is not specific to a single LLM family. LLaMA consistently achieves higher HIT@10 than Qwen across both sequential encoders and all five datasets, with relative differences ranging from 2.2\% to 11.8\%. Importantly, the behavior of the injected collaborative representations remains largely consistent across the two LLMs: SASRec-based representations generally match or outperform BERT4Rec-based representations, while the datasets on which REPREC-B is competitive remain similar across backbones. This consistency suggests that REPREC can transfer collaborative signals across different LLM families without backbone-specific architectural modifications, while the absolute recommendation quality remains influenced by the choice of the underlying LLM.

\subsubsection{\textbf{Effect of LLM Size:}}
\begin{figure}[t]
    \centering
    \includegraphics[width=\columnwidth]{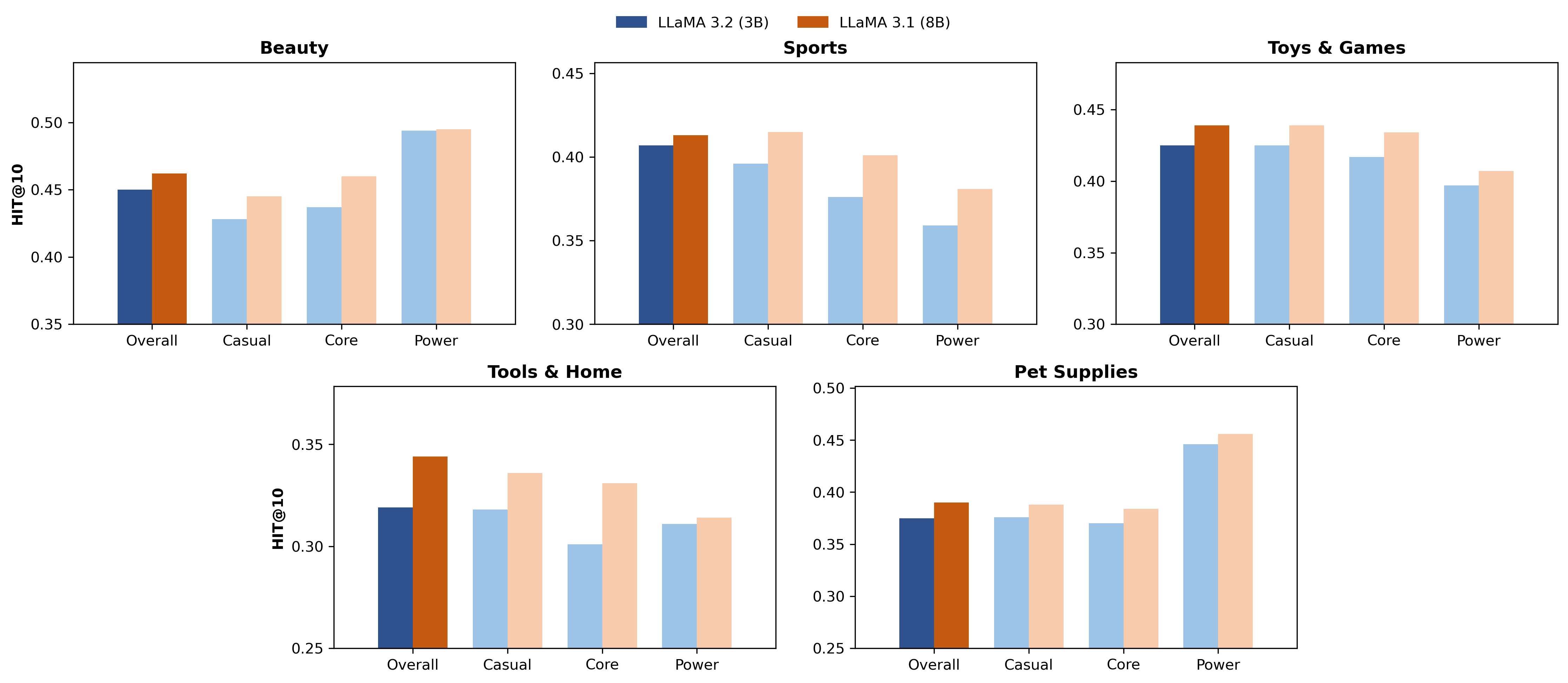}
    \caption{
    Effect of LLM size on REPREC performance across five datasets and user activity groups. 
    }
    \label{fig:llm_size}
\end{figure}

We further investigate how LLM scale affects REPREC by comparing LLaMA-3.2-3B and LLaMA-3.1-8B across datasets and user regimes. As shown in Figure~\ref{fig:llm_size}, increasing the LLM size consistently improves overall HIT@10 across all five datasets, with relative gains ranging from 2.67\% to 7.84\%. When stratified by user activity, the larger LLM improves performance across casual, core, and power users, but the magnitude of improvement differs across regimes. Averaged across datasets, HIT@10 improves by 4.18\% for casual users and 5.95\% for core users, compared with 2.41\% for power users. This suggests that increasing LLM capacity is particularly beneficial when interaction histories are limited, where stronger pretrained knowledge can complement the available collaborative signal. In contrast, the smaller gains for power users suggest diminishing benefits from LLM scaling as richer interaction histories provide stronger personalization signals. Overall, these results show that REPREC continues to benefit from stronger LLM backbones while preserving its effectiveness across user activity levels.

\subsubsection{\textbf{Effect of User-Specific Collaborative Signal:}}
To determine whether REPREC genuinely exploits user-specific collaborative information, rather than treating the injected representations as generic additional conditioning parameters, we introduce two controlled ablations. \textit{Prompt Tuning} replaces the projected user representation $g_\phi(h_u)$ with $m{=}6$ free soft tokens shared across all users, removing the sequential encoder and any collaborative input. \textit{Shuffled} retains the trained REPREC injector but, at inference time, replaces each user's SASRec representation with that of a randomly selected different user. Importantly, the shuffled representation remains a valid collaborative embedding drawn from the same encoder and distribution; only its correspondence with the user's textual interaction history is disrupted. This allows us to test whether any collaborative representation is sufficient as an additional conditioning signal, or whether REPREC actually relies on the user-specific information encoded within it.

As shown in Table~\ref{tab:shuffle-ablation}, REPREC consistently outperforms Prompt Tuning by an average of 11.39\% across datasets, indicating that its gains cannot be explained solely by the additional capacity introduced by soft-token conditioning. More importantly, replacing the matched representation with another user's valid collaborative representation reduces HIT@10 by 11.06\%--30.22\% across all five datasets, with an average degradation of 18.31\%. Since Shuffled preserves the injector, number of soft tokens, and distribution of collaborative representations while changing only which user's representation is provided, this degradation shows that arbitrary collaborative signals are insufficient. Instead, REPREC exploits information encoded in the specific user's sequential representation and aligns it with the textual interaction context. These results provide evidence that the LLM does not simply treat the injected embeddings as generic additional parameters while relying only on the textual prompt; the identity-specific collaborative signal itself contributes meaningfully to recommendation performance.

\begin{table}[t]
\centering
\caption{
Ablation comparing Prompt Tuning, Shuffled, and REPREC. The final column
reports the relative performance drop when replacing the matched user
representation in REPREC with a randomly shuffled user representation.
}
\label{tab:shuffle-ablation}

\scriptsize
\setlength{\tabcolsep}{3.5pt}
\renewcommand{\arraystretch}{1.08}

\resizebox{\columnwidth}{!}{%
\begin{tabular}{lccc|c}
\toprule

\textbf{Dataset}
& \textbf{Prompt Tuning}
& \textbf{Shuffled}
& \textbf{REPREC}
& \textbf{REPREC $\rightarrow$ Shuffled} \\

\midrule

Beauty
& 0.394
& 0.356
& \textbf{0.450}
& \textcolor{red}{\(\downarrow\)} 20.89\% \\

Sports
& 0.366
& 0.284
& \textbf{0.407}
& \textcolor{red}{\(\downarrow\)} 30.22\% \\

Toys \& Games
& 0.386
& 0.378
& \textbf{0.425}
& \textcolor{red}{\(\downarrow\)} 11.06\% \\

Pet Supplies
& 0.345
& 0.313
& \textbf{0.375}
& \textcolor{red}{\(\downarrow\)} 16.53\% \\

Tools \& Home
& 0.283
& 0.278
& \textbf{0.319}
& \textcolor{red}{\(\downarrow\)}12.85\% \\

\bottomrule
\end{tabular}%
}
\end{table}

\section{Conclusion}
This work demonstrates that collaborative signals can be effectively integrated into LLM-based recommendation through lightweight user-level representation alignment. REPREC maps representations from a frozen sequential encoder into a small number of soft tokens that condition a frozen LLM, avoiding LLM fine-tuning and item-level representation injection. Across five datasets, two sequential encoders, multiple LLM backbones, and different LLM sizes, REPREC consistently achieves strong performance against sequential and LLM-based baselines, demonstrating its effectiveness across model configurations. Its gains extend across casual, core, and power users, showing that compact collaborative representations remain valuable across varying levels of interaction history. Beyond recommendation quality, REPREC is parameter, training, and inference-efficient: it requires neither LLM fine-tuning nor multi-stage training and avoids the token overhead associated with item-level representation injection. Furthermore, training with shorter histories and leveraging longer histories at inference reduces training cost while preserving most of the full-history performance, providing a practical efficiency-performance trade-off. Overall, REPREC offers a simple, modular, and computationally efficient framework for incorporating user-specific collaborative knowledge into LLM-based recommendation.

\section{Future Work}
For future work we want to extend REPREC to a broader range of sequential encoders, such as GRU4Rec and MAMBA. We also plan to evaluate REPREC on datasets with substantially longer interaction histories to further investigate its ability to train with shorter prompt histories while effectively leveraging longer contexts at inference time.

\section{Ethical Considerations}

Our experiments are conducted on established Amazon product review benchmarks and use historical interaction data for next-item recommendation. We do not introduce additional sources of personal or sensitive information beyond the interaction data provided by these benchmarks. Nevertheless, user representations learned from behavioral histories may encode information about individual preferences and should therefore be handled carefully in real-world deployments. Appropriate data-governance, privacy, and access-control mechanisms would be necessary when applying REPREC to proprietary user data.

Like other recommender systems, REPREC may inherit or amplify biases present in the underlying interaction data, including popularity and exposure biases. Our evaluation focuses primarily on recommendation accuracy and computational efficiency and does not explicitly measure fairness across demographic groups or other socially relevant dimensions. Consequently, strong recommendation performance should not be interpreted as evidence of fairness or absence of bias. Future work could examine fairness, privacy-preserving representation learning, and robustness across different user populations.

Finally, REPREC is intended as a general framework for efficient recommendation. Its deployment in high-stakes or sensitive domains would require additional evaluation of domain-specific risks, potential harms, and appropriate human oversight beyond the benchmark experiments considered in this work.

\bibliographystyle{ACM-Reference-Format}
\bibliography{sample-base}



\appendix

\section{Parameter Count Analysis}
\label{app:params}
\subsection{LoRA Trainable Parameters}
\label{app:params:lora}
We apply LoRA to the query (\texttt{q\_proj}) and value
(\texttt{v\_proj}) projection matrices of every transformer
layer in the frozen LLaMA-3.2-3B-Instruct backbone
($L{=}28$ layers, hidden size $D{=}3072$, 24 attention heads,
8 key-value heads via Grouped Query Attention).
Each low-rank adapter introduces two factor matrices per
target projection, giving a per-rank scaling of
$286{,}720$ parameters: $|\theta_{\text{LoRA}}| \;=\; 286{,}720 \times r.$ Table~\ref{tab:parameter_counts} reports the resulting counts for
$r \in \{1, 4, 8, 16, 32\}$.

\subsection{REPREC Trainable Parameters}
\label{app:params:reprec}
The sequential encoder (SASRec) is pre-trained and frozen
prior to injector training; only the injector MLP parameters
are updated end-to-end. The injector maps the frozen SASRec user embedding
$\mathbf{u} \in \mathbb{R}^d$ to $m$ soft conditioning tokens
in the LLM embedding space: $
g_\phi(\mathbf{u}) =
\mathrm{LN}_{D}\!\left(
  \operatorname{reshape}_{m \times D}\!\left(
    \mathbf{W}_2\,\sigma(\mathbf{W}_1\mathbf{u} + \mathbf{b}_1) + \mathbf{b}_2
  \right)
\right)
$, where $\mathbf{W}_1 \in \mathbb{R}^{h_d \times d}$,
$\mathbf{W}_2 \in \mathbb{R}^{(m \cdot D) \times h_d}$,
$h_d{=}128$ is the intermediate hidden size, $D{=}3072$ is
the LLM hidden size, and $\mathrm{LN}$ is a LayerNorm applied
over the $D$-dimensional soft token embeddings after reshape.
The total parameter count is:

\begin{equation}
  |\theta_{\text{inj}}|
  = \underbrace{(d \cdot h_d + h_d)}_{\text{Layer 1}}
  + \underbrace{(h_d \cdot mD + mD)}_{\text{Layer 2}}
  + \underbrace{2D}_{\text{LayerNorm}}
  \label{eq:injector_params}
\end{equation}

\begin{table*}[t]
  \centering
  \caption{Trainable-parameter counts for LoRA and the REPREC injector
  under different configurations. Rows marked $\dagger$ denote the
  configurations used in the main experiments.}
  \label{tab:parameter_counts}
  \small

  \begin{minipage}[t]{0.54\textwidth}
    \centering
    \textbf{(a) LoRA trainable parameters}

    \vspace{0.4em}

    \resizebox{\linewidth}{!}{%
    \begin{tabular}{lrr}
      \toprule
      \textbf{Rank} $r$
        & \textbf{Params / layer}
        & \textbf{Total params} \\
      \midrule
      1  & 10,240  & 286,720 \\
      4  & 40,960  & 1,146,880 \\
      \rowcolor{gray!15}
      8$^\dagger$ & 81,920 & 2,293,760 \\
      16 & 163,840 & 4,587,520 \\
      32 & 327,680 & 9,175,040 \\
      \bottomrule
    \end{tabular}%
    }

    \vspace{0.4em}

    \footnotesize
    \texttt{q\_proj} and \texttt{v\_proj} only, using
    LLaMA-3.2-3B-Instruct with $D{=}3072$, $L{=}28$, and
    GQA with 8 KV heads.
  \end{minipage}
  \hfill
  \begin{minipage}[t]{0.42\textwidth}
    \centering
    \textbf{(b) REPREC injector parameters}

    \vspace{0.4em}

    \resizebox{\linewidth}{!}{%
    \begin{tabular}{lr}
      \toprule
      \textbf{Soft tokens} $m$
        & \textbf{Trainable params} \\
      \midrule
      2 & 807,040 \\
      4 & 1,599,616 \\
      \rowcolor{gray!15}
      6$^\dagger$ & 2,392,192 \\
      8 & 3,184,768 \\
      \bottomrule
    \end{tabular}%
    }

    \vspace{0.4em}

    \footnotesize
    Injector dimensions are $d{=}64$, $h_d{=}128$, and
    $D{=}3072$. SASRec remains frozen. The $m{=}6$
    configuration approximately matches LoRA $r{=}8$
    (2.39M versus 2.29M parameters).
  \end{minipage}
\end{table*}

\section{Negative Sampling Strategy}
\label{app:negsampling}
 
To ensure reproducible and fair evaluation across all methods,
we fix the negative item sets prior to training using a hybrid
\textbf{Random + Popularity-based Negative Sampling} (PNS)
strategy.
For each test user, we sample $K{=}200$ negative items
from the full item pool.
Negatives are drawn from the \emph{entire} item catalogue
without excluding user-interacted items, following standard
practice in sequential recommendation evaluation.
 
\paragraph{Hybrid sampling procedure.}
Let $\mathcal{V}$ denote the full item vocabulary with
$|\mathcal{V}|$ items.
For each item $i \in \mathcal{V}$, let $c_i$ be its
interaction count in the training split.
We define a popularity weight with Laplace smoothing:
 
\begin{equation}
  w_i = c_i + 1, \quad i \in \mathcal{V},
  \label{eq:pop_weight}
\end{equation}
 
\noindent so that every item (including unseen ones) has
non-zero sampling probability.
Given a popularity ratio $\rho \in [0,1]$, for each test
user $u$ we draw:
 
\begin{itemize}
  \item $K_{\text{pop}} = \lfloor \rho K \rfloor$: negatives
        sampled proportionally to item popularity weights
        $\{w_i\}$.
  \item $K_{\text{rand}} = K - K_{\text{pop}}$: negatives
        sampled uniformly at random from $\mathcal{V}$.
\end{itemize}
 
\noindent The two sets are concatenated and shuffled.
We use $\rho{=}0.5$, yielding 100 popularity-weighted and
100 uniform-random negatives per user.
Popularity-weighted negatives serve as informative hard negatives: items with high interaction frequency are plausible candidates for any user — regardless of whether the target user has personally interacted with them — making them more discriminative than rare items and providing a more realistic evaluation signal than pure random sampling.

\section{Training and Validation Protocol}
\label{app:training_protocol}

\subsection{Training Objective}
\label{app:training_objective}

REPREC is trained with the standard causal language-modeling
objective. The input prompt is masked out of the loss, and
gradients are computed only over the target answer token
(\texttt{Yes} or \texttt{No}):

\begin{equation}
\mathcal{L}_{\mathrm{CLM}}
=
-\log P_{\phi}
\left(
y_t \mid x_{1:t-1}
\right),
\label{eq:clm_objective}
\end{equation}

where $y_t \in \{\texttt{Yes},\texttt{No}\}$ is the answer token and
$x_{1:t-1}$ is the conditioned prompt. This objective is aligned
with inference, where the recommendation score is obtained from
$\log P(\texttt{Yes})$ at the same answer position — the training
signal and the inference score are derived from the same
conditional token distribution.

Unlike methods that substitute conditioning information at
placeholder token positions within the prompt text, REPREC's
injector module maps the frozen SASRec user embedding
$\mathbf{u}\in\mathbb{R}^d$ to $m$ soft conditioning tokens
$g_\phi(\mathbf{u}) \in \mathbb{R}^{m\times D}$(Appendix~\ref{app:params:reprec}),
which are \emph{prepended} to the token embedding sequence before
the text prompt is consumed by the LLM. The text prompt itself is
therefore unmodified and identical for every candidate item:

\begin{quote}
\ttfamily
User has the following purchase history:\\
\{item$_1$\}, \{item$_2$\}, \ldots, \{item$_n$\}\\
Is the user likely to buy \{candidate\} next?
\end{quote}

with the $m$ soft tokens occupying the $m$ positions immediately
preceding this text at the embedding level. At inference, we obtain next-token logits from the final non-padding prompt position and use the log-probability assigned to the Yes token as the candidate relevance score (Eq.~\ref{eq:clm_objective}).

\subsection{Training-Time Negative Sampling}
\label{app:negative_sampling}

Following standard practice in sequential
recommendation, each positive training interaction
is paired with $K_{\mathrm{train}}=2$ sampled negative items
(\texttt{neg\_per\_pos}=2), rather than using the full candidate pool
employed during evaluation. In contrast, ranking
metrics at evaluation time require the target item to be ranked
against a substantially larger candidate set.

\subsection{Training-Time Quick Validation}
\label{app:quick_validation}

Performing the complete evaluation protocol after every training
epoch is computationally expensive because it requires repeatedly
scoring every validation user against the full candidate pool.
We therefore use a reduced validation pass solely for model
selection during training. At the start of training, a fixed
subsample of
\[
N_{\mathrm{val}} = 5{,}048
\]
users is drawn (with a fixed seed) from the full validation set and
reused for every epoch's quick-validation pass. Each sampled user is
scored against a reduced pool of
\[
K_{\mathrm{val}} = 100
\]
negative items drawn from the same fixed negative pool used for
final evaluation.

\section{Baseline Implementation}

\subsection{\textbf{TALLRec:}} We do not train TALLRec as a separate model.
TALLRec's original recipe consists of (i) an Alpaca-style
instruction-tuning warm-start on a non-instruction-tuned LLaMA-1
backbone, followed by (ii) LoRA fine-tuning (rank~8, target
modules \texttt{q\_proj}/ \texttt{v\_proj}) on a preference/
unpreference-framed prompt, typically under a few-shot training
regime. In our setting, none of these three properties hold: the
backbone is already instruction-tuned
(LLaMA-3.2-3B-\textit{Instruct}), training uses the full dataset
rather than a few-shot subset, and the preference/unpreference
framing requires explicit negative (disliked) feedback, which our
implicit purchase-history data does not provide. Under these
conditions, TALLRec's LoRA configuration and resulting text-only
conditioning are identical to LoRA-FT's. We therefore report
TALLRec's performance via LoRA-FT rather than training it as a
separate model.

\subsection{\textbf{LLaRA:}} The frozen sequential encoder's raw per-item
embedding table is projected
into LLM token space by a trainable two-layer MLP. Each history
item and the candidate item are followed in the prompt by an
\texttt{[ItemEmb]} placeholder token, whose embedding is replaced
in place with the corresponding projected item embedding before the
sequence is consumed by the LLM. The LoRA adapter and the projector
are trained jointly, single-stage. We use LLaRA's "Direct Training" configuration (no curriculum prompt tuning): the hybrid textual+behavioral representation is used throughout training.

\subsection{\textbf{CoLLM:}} We follow the original two-stage recipe. Stage 1
fine-tunes LoRA on text-only prompts — equivalent to, and in our
implementation reusing, a LoRA-FT checkpoint. Stage 2 freezes that
adapter and trains only a projector: a shared two-layer MLP that
maps both a pooled user embedding and a candidate-item embedding
(both from the frozen sequential encoder) into LLM token space. The
prompt inserts a single \texttt{[UserID]} placeholder at the user
mention and a single \texttt{[TargetItemID]} placeholder at the
candidate mention; both are substituted in place with the
projector's output. Stage 2 is trained with the binary
cross-entropy objective, following the
original method rather than the causal-LM objective used by the
other baselines.

\end{document}